\documentclass[prb,aps,superscriptaddress,reprint]{revtex4-1}

\usepackage[english]{babel}
\usepackage{graphicx}
\usepackage{bm}
\usepackage[utf8]{inputenc}
\usepackage[T1]{fontenc}
\usepackage{mathptmx}
\usepackage{SIunits}
\usepackage{color} 
\usepackage[colorlinks=true,urlcolor=blue,linkcolor=black,citecolor=black,]{hyperref}

\begin{document}

\title{How to solve problems in micro- and nanofabrication caused by the emission of electrons and charged metal atoms during e-beam evaporation}

\author{Frank Volmer}
\affiliation{2nd Institute of Physics and JARA-FIT, RWTH Aachen University, 52074 Aachen, Germany}

\author{Inga Seidler}
\affiliation{JARA-FIT Institute for Quantum Information, Forschungszentrum J\"ulich GmbH and RWTH Aachen University, Aachen, Germany}

\author{Timo Bisswanger}
\affiliation{2nd Institute of Physics and JARA-FIT, RWTH Aachen University, 52074 Aachen, Germany}

\author{Jhih-Sian Tu}
\affiliation{Helmholtz Nano Facility (HNF), Forschungszentrum J\"ulich, 52425 J\"ulich, Germany}

\author{Lars R. Schreiber}
\affiliation{JARA-FIT Institute for Quantum Information, Forschungszentrum J\"ulich GmbH and RWTH Aachen University, Aachen, Germany}

\author{Christoph Stampfer}
\affiliation{2nd Institute of Physics and JARA-FIT, RWTH Aachen University, 52074 Aachen, Germany}
\affiliation{Peter Gr\"unberg Institute (PGI-9), Forschungszentrum J\"ulich, 52425 J\"ulich, Germany}

\author{Bernd Beschoten}
\affiliation{2nd Institute of Physics and JARA-FIT, RWTH Aachen University, 52074 Aachen, Germany}

\email{E-mail address: bernd.beschoten@physik.rwth-aachen.de}

\begin{abstract}
We discuss how the emission of electrons and ions during electron-beam-induced physical vapor deposition can cause problems in micro- and nanofabrication processes. After giving a short overview of different types of radiation emitted from an electron-beam (e-beam) evaporator and how the amount of radiation depends on different deposition parameters and conditions, we highlight two phenomena in more detail: First, we discuss an unintentional shadow evaporation beneath the undercut of a resist layer caused by the one part of the metal vapor which got ionized by electron-impact ionization. These ions first lead to an unintentional build-up of charges on the sample, which in turn results in an electrostatic deflection of subsequently incoming ionized metal atoms towards the undercut of the resist. Second, we show how low-energy secondary electrons during the metallization process can cause cross-linking, blisters, and bubbles in the respective resist layer used for defining micro- and nanostructures in an e-beam lithography process. After the metal deposition, the cross-linked resist may lead to significant problems in the lift-off process and causes leftover residues on the device. We provide a troubleshooting guide on how to minimize these effects, which e.g. includes the correct alignment of the e-beam, the avoidance of contaminations in the crucible and, most importantly, the installation of deflector electrodes within the evaporation chamber. 
\end{abstract}

\maketitle

\section{Introduction}
Electron-beam-induced physical vapor deposition (e-beam deposition) has developed into a versatile tool in the field of micro- and nanofabrication:\cite{Harsha2006, Ohring2002, Mattox2010} Compared to resistive evaporation techniques, e-beam deposition allows the evaporation of a larger variety of materials, including many oxides and metals with very low vapor pressures. Furthermore, the direct heating of the evaporation material by the electron beam allows the crucible to be water-cooled. This minimizes reactions between the molten metal and the crucible material. Compared to, e.g., sputtering techniques, e-beam deposition allows directional growth conditions, the avoidance of plasma-induced defects, and minimizes the incorporation of residual gases into the deposited layer.

But there are also drawbacks in using e-beam deposition, which include the generation of X-rays, the emission of electrons over a large energy range, and the creation of ions by electron-impact ionization, which all can lead to problems in device fabrication.\cite{Harsha2006, Ohring2002, Mattox2010} Especially in the field of semiconductor technology, it is well documented that radiation emitted by an e-beam evaporator can induce defects in the semiconductor material.\cite{Mayo1976, El-Kareh1978, Auret1984, Blakers1984, Auret2006} Possible side-effects of e-beam evaporation on other aspects of micro- and nanofabrication, e.g., lift-off problems of resist defined patterns,\cite{Pao1999, Cheng2010} are far less reported. Therefore, in this article we first give an overview on different types of radiation that are emitted during e-beam evaporation and how the amount of radiation depends on different deposition parameters and conditions (section \ref{Overview}). Then, we discuss an unintentional shadow evaporation beneath the undercut of a resist layer caused by the partially charged metal vapor in section \ref{Unintentional_Shadow_Deposition}. In section \ref{Damage_Resist_Lift-Off_Problems} we show how low-energy secondary electrons can cause damage to an e-beam resist based on polymethyl methacrylate (PMMA). The interaction between resist and electrons leads to both cross-linking and bubbles, which results in significant issues during lift-off. Finally, we give a troubleshooting guide on how to minimize these side-effects of e-beam evaporation in section \ref{Troubleshooting}. 

\section{Types of particles and radiation emitted from an e-beam evaporator}
\label{Overview}
\subsection{Partially ionized vapor}
\label{Partially_ionized_vapor}
\begin{figure}[tb]
	\includegraphics{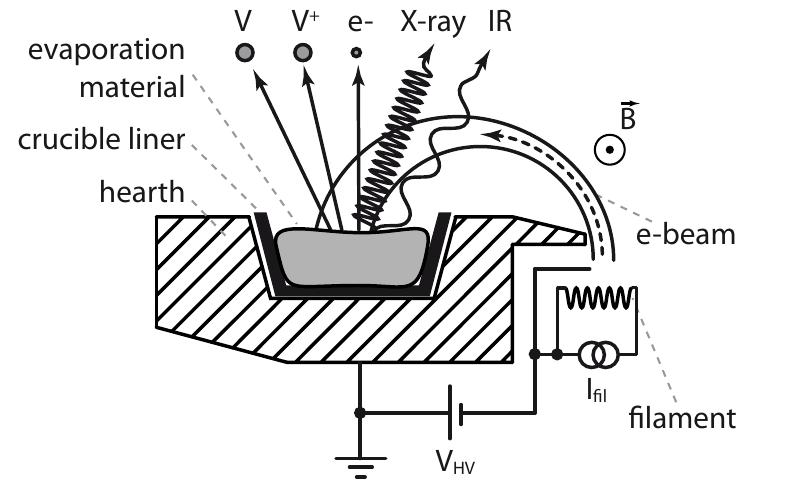}
	\caption{In an e-beam evaporator a hot filament emits electrons due to thermionic emission. These electrons are accelerated by a dc voltage $V_\text{HV}$ and focused onto the surface of the evaporation material by a magnetic field $\vec{B}$. A part of the vapor of the evaporation material (V) will be ionized by electron-impact ionization (V$^+$) as the vapor has to cross the same region where the e-beam travels. Next to infrared radiation (IR) generated by the hot material, the e-beam will also create bremsstrahlung (X-rays) and secondary electrons (e$^-$).}
	\label{Fig1}
\end{figure}

The most obvious particles, which get emitted from an e-beam evaporator, is the actual evaporation material in its vapor phase (see Fig.~\ref{Fig1} where we denote the vapor in its neutral form as 'V'). To estimate the density of this vapor at low deposition rates, we can assume molecular flow conditions and calculate the vapor pressure by the Hertz-Knudsen equation (also called Hertz-Langmuir-Knudsen equation).\cite{Harsha2006, Ohring2002, Mattox2010} For higher evaporation rates, the density of the vapor just above the crucible is high enough that collisions between particles have to be taken into account. This results in a viscous cloud of evaporated material, called virtual source, which will change the spatial distribution of the vapor beam.\cite{Graper1973,Chaleix1996}

But it is exactly this region of high vapor density that the e-beam has to cross on its way to the crucible (see Fig.~\ref{Fig1}). Therefore, electron-impact ionization between primary electrons and the vapor leads to the partial ionization of the vapor.\cite{Graper1970,Nishio1992,Nishio1993,Besuelle1998,Bhatia1998,Majumder2010,Mukherjee2016} Several factors such as the alignment of the e-beam, chosen deposition parameters, and material constants of the evaporation material determine the total amount of this ionization, i.e. the ratio between neutral and ionized vapor densities.\cite{Bhatia1998,Mukherjee2016,Gasper2015} Especially at higher deposition rates, when a virtual source is created slightly above the actual crucible,\cite{Graper1973,Chaleix1996} the amount of ionization increases, if the focus of the e-beam moves into this viscous cloud.\cite{Bhatia1998}

The insertion of a crucible liner can reduce the amount of ionization. As the liner includes a thermal resistance between evaporation material and water cooled hearth, the heat loss will be reduced and the same evaporation rate can be achieved with less power,\cite{Harsha2006, Ohring2002, Mattox2010} i.e., less e-beam current and therefore less total electron-impact ionization. At the same time, the thermal resistance of the liner also reduces the temperature gradient between the heat source, i.e. the area at which the e-beam is hitting the crucible, and the rest of the crucible. This is argued to increase the area at which evaporation occurs compared to the area of direct electron beam bombardment and, hence, decreases the chance of electron-impact ionization further.\cite{Mukherjee2016}

So far, the discussion about the amount of ionization has focused on the interaction of the vapor with the primary electrons of the e-beam. In addition, backscattered and secondary electrons, which are created by the interaction of the e-beam with the evaporation material, are important contributors to the overall ionization process, which will be discussed in the next subsection.\cite{Nishio1992,Dikshit2005}

\subsection{Secondary and backscattered electrons}
\label{Secondary_and_backscattered_electrons}
The interaction between the primary electrons of the e-beam and the evaporation material generates a magnitude of different types of electrons with different energies (denoted as 'e$^-$' in Fig.~\ref{Fig1}): Elastically reflected electrons, inelastically backscattered electrons, secondary electrons, electrons created during electron-impact ionization, electrons from thermal ionization, and finally thermionic electrons.\cite{Seiler1983,Nishio1992} A part of these electrons can reach the substrate during the evaporation process, which can lead to defects in semiconductors\cite{Myburg1992, Auret2006} or problems with a resist layer.\cite{Pao1999, Cheng2010}. The energy distribution and yield of elastically reflected and inelastically backscattered electrons depend on the atomic number of the evaporation material and the electron primary energy (i.e. the applied acceleration voltage).\cite{Sternglass1954,Archard1961} It was also demonstrated that the spatial distribution of reflected electrons depends on the alignment of the e-beam.\cite{Yamada2011} 

Electrons emitted from the evaporation material are called secondary electrons (SE). Their kinetic energies are usually less than \unit{50}{eV}.\cite{Seiler1983,Nishio1992} For the vapor ionization process discussed in section \ref{Partially_ionized_vapor}, the high-energy tail of the SEs' energy distribution gets important, as the cross section of the electron-impact ionization exhibits a maximum in the energy range of \unit{10-50}{eV} for typical materials.\cite{Freund1990,Bartlett2002}

The average number of SE, which are emitted per primary electron, is called secondary electron yield (SEY). The SEY is found to depend on the angle of incidence of the primary electrons,\cite{Seiler1983,Thomas1969,Kanaya1972} and therefore on the adjustment of the e-beam. This is due to the fact that only SE excited within a certain depth of the evaporation material can reach and subsequently escape the surface. The smaller the angle between the e-beam and the surface, the longer the penetration distance of the e-beam within this escape depth and, hence, the higher the SEY. The escape depth can vary between different materials and can become very high for insulators.\cite{Seiler1983} The SEY also depends on prior treatments and on the morphology of the surface which gets hit by the e-beam.\cite{Millet1995,Baglin2000,Dikshit2005} Crucially, the SEY can increase in the presence of contaminants, an oxide layer, or adsorbates which cover the evaporation material.\cite{Gonzalez2017,Baglin2000}

\subsection{Electromagnetic radiation}
\label{Electromagnetic_radiation}
The thermal radiation from the hot evaporation material is one source of electromagnetic radiation.\cite{Harsha2006, Ohring2002, Mattox2010} For typical evaporation temperatures, the vast majority of emitted photons have energies within the infrared range (denoted IR in Fig.~\ref{Fig1}) according to Planck's law. The temperature of the evaporation material can be calculated once the vapor pressure for a given deposition rate and chamber geometry is estimated by the Hertz-Knudsen equation. For this, the dependence between vapor pressure and temperature can be deduced from the Clausius-Clapeyron equation as:\cite{Harsha2006, Ohring2002, Mattox2010} 
\begin{equation}
 \log_{10}(p)=-\frac{A}{T}+B\log_{10}(T)+CT+DT^{-2}+E,
\end{equation}
where $p$ is the vapor pressure, $T$ is the temperature and $A$ to $E$ are material-specific coefficients. The coefficients $B$ to $D$ normally only contribute smaller corrections to the other terms and, hence, the simplified Antoine equation ($\log _{10}p=-A/T+E$) describes the temperature dependent vapor pressure in most cases sufficiently well.\cite{Honig1969}

\begin{figure*}[tb]
	\includegraphics{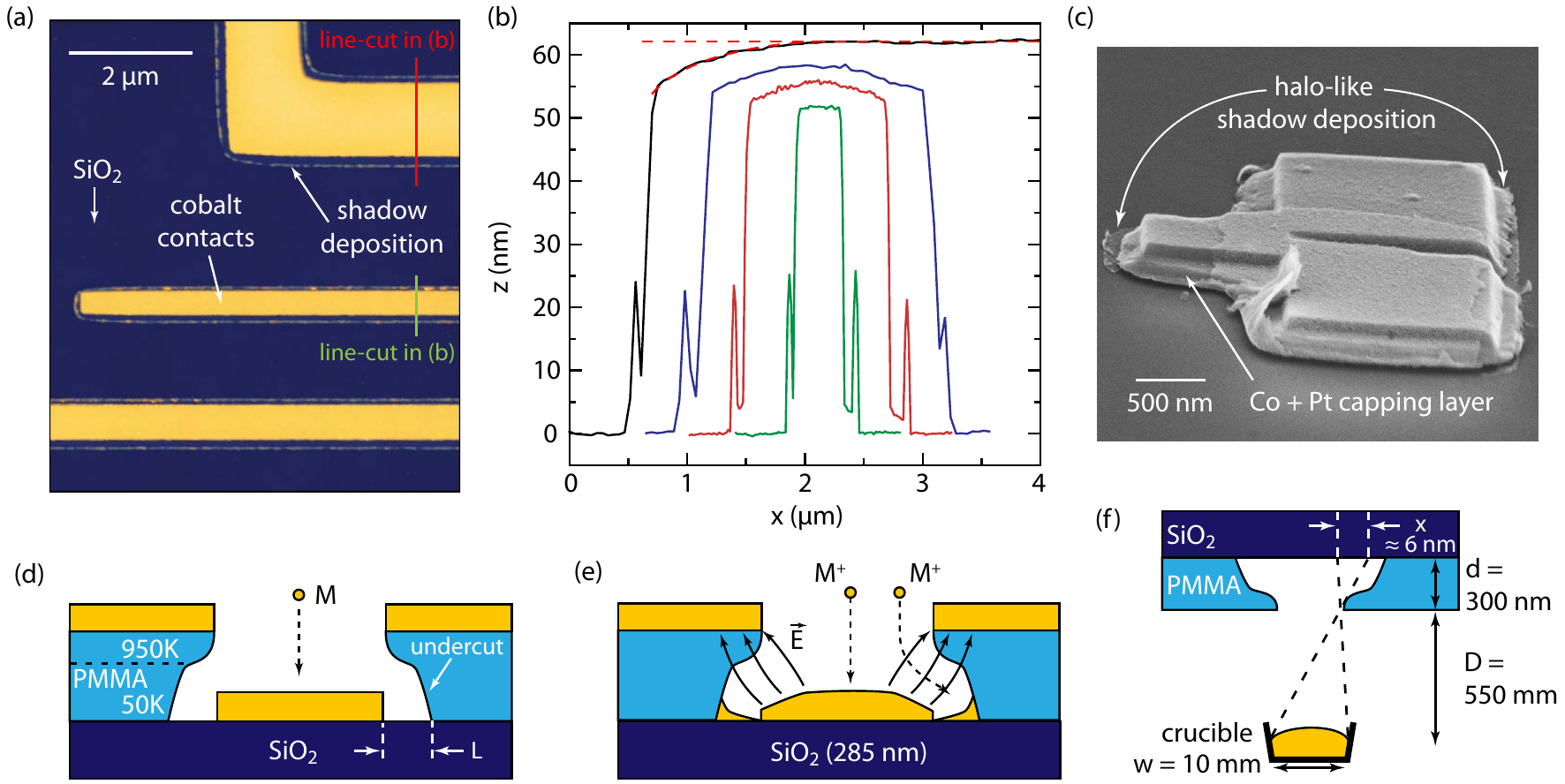}
	\caption{(a) Scanning force microscopy (SFM) image of cobalt contact electrodes deposited onto a Si$^{++}$/SiO$_2$ substrate and (b) line-cuts both extracted from this and other SFM images. Along the edges of the contacts a shadow deposition can be observed as fence-like structures. The line-cuts reveal that only contacts with sufficiently large widths exhibit flat surfaces approximately $\unit{1}{\micro m}$ away from an edge (black curve in (b), dashed lines are guides to the eye). Due to the curved edges, the height at the center of the contacts decreases as soon as the curvatures from opposite edges start to converge. (c) Scanning electron microscopy (SEM) image of a cobalt nanostructure capped with platinum showing a less pronounced occurrence of the shadow evaporation which results in a halo-like structure. (d) An ideal metalization scheme of a prepatterned nanostructure written in a double resist layer with an undercut. (e) If the metal vapor is partially ionized, the deposited metal layer on the insulating substrate can have a different electrical potential compared to the metal layer on top of the resist due to a different overall resistance to ground. The resulting electrostatic deflection of incoming charged metal atoms will lead to the unwanted shadow deposition in the undercut region of the resist layer and the curved edges of the deposited structure. (f) The shadow deposition has a much larger spatial dimension than expected from normal blurring caused by the projection of the crucible on the edge of the undercut.}
	\label{Fig2}
\end{figure*}

The exponential dependence of the vapor pressure as a function of temperature has an important implication: At around a temperature at which the evaporation starts, there will be a significant increase in vapor pressure and hence in the deposition rate for only a small further increase in temperature. At the same time, the thermal radiation from the crucible, which scales polynomially with $T^4$ according to the Stefan-Boltzmann law, increases much less. Therefore, for a given deposition time the ratio of the amount of deposited material to the amount of transferred heat by thermal radiation decreases with increasing temperature. Infrared radiation is not the only source of heat in an evaporation process: At high deposition rates, the heat from condensation (equivalent to the heat of vaporization), which can be up to a few eV per atom, can dominate the overall heat transfer to the substrate.\cite{Pargellis1989,Harsha2006}

Another source of electromagnetic radiation during e-beam evaporation is bremsstrahlung, which is created by the impact of primary electrons onto the evaporation material.\cite{Mayo1976, Bhattacharya1988} As acceleration voltages between 4 and \unit{20}{kV} are typically used in e-beam evaporators,\cite{Harsha2006, Ohring2002, Mattox2010} the spectral density of the bremsstrahlung has its maximum within the X-ray regime. These high energy photons can be an important source of damage to a device.\cite{Mayo1976,Hordequin2001,Yamada2011,Rybicki2012} To reduce the total dose of X-ray radiation, we argue similarly as for the thermal radiation: In a first order approximation, the temperature of the evaporation material increases linearly with e-beam current for a fixed acceleration voltage. According to the Antoine equation, this can lead to an exponentially increasing deposition rate as a function of e-beam current. A slight increase in e-beam current can therefore dramatically reduce the overall deposition time for a given thickness. On the other hand, the X-ray flux only scales linearly with the e-beam current.\cite{Bhattacharya1988,Yamada2011} Therefore, the X-ray dose integrated over growth time will decrease with both higher deposition rates and e-beam currents.\cite{Yamada2011} It is important to note that this argumentation holds for every other kind of radiation or particles where the respective flux scales approximately linear with e-beam current, e.g. the flux of secondary and backscattered electrons.

\section{Shadow deposition due to ionized metal vapor}
\label{Unintentional_Shadow_Deposition}
Shadow evaporation beneath the undercut of a resist layer is one effect of charges emitted from an e-beam evaporator. This shadow evaporation can lead to a fence-like structure observed along the edges of the structures in the scanning force microscopy (SFM) image in Fig.~2(a). The depicted structures are cobalt contact electrodes (in the following called contacts) used in spintronic devices.\cite{Volmer2014,Drogeler2014Nov,Volmer2015Apr,Droegeler2016} They are fabricated by first spin coating a double resist layer on top of a Si$^{++}$/SiO$_2$ (\unit{285}{nm}) substrate. This double layer consists of a 50K and a 950K PMMA layer and is used to create an undercut resist profile in the e-beam lithography and subsequent development (see Fig.~2(d)).\cite{Cui2008} Under directional growth conditions such undercuts should prevent the deposition of evaporation material on the resist's sidewalls and therefore should lead to well-defined structures as sketched in Fig.~2(d). This simple model, however, is disproved by the fence-like structure observed in Fig.~2(a). Since the fence-like structures appear regardless of the orientation of the contacts, they cannot be explained by a simple misalignment of the sample surface from the normal incidence of the deposition direction. It is reasonable to assume a directional molecular vapor beam as we used a low deposition rate of $\unit{0.02}{nm/s}$ and the chamber pressure during the evaporation was in the lower $\unit{10^{-9}}{mbar}$ range. Furthermore, we observe that the distance between the edge of the contact and the fence-like structure scales with the length $L$ of the undercut (see Fig.~2(d)). This length can be influenced by the acceleration voltage during the e-beam lithography process.\cite{Kyser1975,Cui2008} For the device shown in Fig.~2(a), a relatively small voltage of \unit{10}{keV} was used, yielding a large undercut. For higher acceleration voltages, the fence-like structure approaches the contact.

Figure 2(b) depicts SFM line-cuts across contacts of various widths (see lines in Fig.~2(a)), revealing a curvature of the contacts' surfaces. Only for sufficiently wide structures, the curvature flattens and converts to a plateau approximately \unit{1}{\micro m} away from the electrode's edge (see black curve in Fig.~2(b), dashed red lines are guides to the eye). If the electrode's width is below $\unit{2}{\micro m}$, the curvatures from opposite sides start to converge, leading to a decreasing overall height. Overall, these line-cuts suggest that the material of the fence-like structures is part of the missing material responsible for the curving of the surface near the edges.

These findings can be explained by a buildup of charges on the deposited metal layer by the partially ionized vapor from the e-beam evaporator. This buildup of charges is possible as the structures in Fig.~2(a) are deposited on a \unit{285}{nm} thick insulating SiO$_2$ layer. In this case, the metal layer on top of the resist can be on a different electrical potential compared to the metal deposited directly on top of the SiO$_2$ due to an overall different resistance to ground. This results in an electrostatic field $\vec{E}$, which is especially pronounced at the edges of the deposited structure (see Fig.~2(e)). Depending on the trajectory of the incoming charged metal atoms, they will either be barely influenced by the electric fields (plateau-like center) or the ions will be deflected towards the undercut region instead of being deposited at the edges of the contacts. For smaller electrode widths the distance between the center of the electrode and the metal layer on top of the resist layer decreases, leading to a situation where also the charged metal atoms with an initial trajectory towards the center of the contact will experience large enough electric fields for a significant electrostatic deflection.

\begin{figure*}[tb]
	\includegraphics{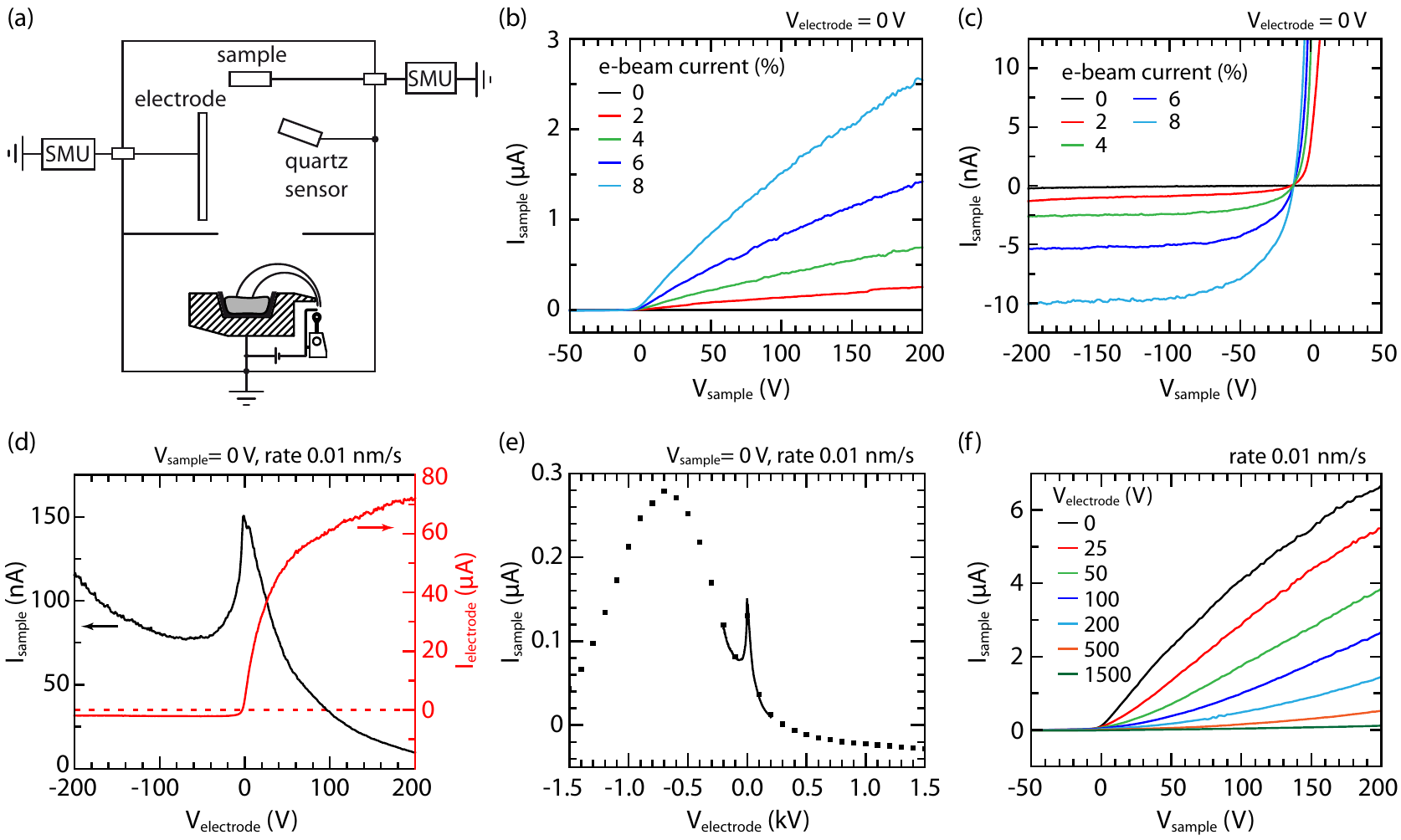}
	\caption{(a) Schematic layout of the evaporation system with an electrode between the crucible and the sample holder with the substrate (whole assembly is called 'sample') to prevent the ionized part of the metal vapor from reaching the sample. Both the sample and the electrode are connected via feedthroughs to source-meter units (SMU). (b) The current flowing through the sample vs. the applied voltage to the sample for different values of e-beam current in case of a grounded electrode. (c) Zoom-in of (b) for negative voltages applied to the sample. (d) Current flowing over both the sample and the electrode for varying voltages applied to the electrode in case of a grounded sample and a constant deposition rate. For high enough positive voltages, the electrode extracts the vast majority of electrons, resulting in an almost vanishing current over the sample. (e) Current over the sample as in (d) but for higher potentials applied to the electrode. (f) Current over the sample vs. the voltage applied to the sample for different potentials of the electrode in case of a constant deposition rate.}
	\label{Fig3}
\end{figure*}

An electrostatic deflection therefore explains the several hundreds of nm wide curvature seen at the edges of the nanostructure in Fig.~2(b), which is much larger than a possible curvature due to a simple geometrical effect: Since the source of the vapor beam, i.e. the crucible, is not a point-like source but has a finite size $w$, the shadow which is cast by the undercut is blurred to some extent (see Fig.~2(f)). Using the mathematical intercept theorem one can estimate the spatial extent $x$ of this blurring in case of the sample shown in Fig.~2(a) to $x=(d/D)\cdot w\approx\unit{6}{nm}$, with $d$ the thickness of the resist layer and $D$ the distance between substrate and crucible. The value of this blurring effect is two orders of magnitudes smaller than the observed curvature of the contacts. We note that the explanation of an electrostatic deflection of incoming ions also has great resemblances to a phenomena known in reactive ion etching (RIE), called either notching or footing.\cite{Hwang1997,Kim2004b,Kim2004}

Further below, we demonstrate that the ionized vapor is indeed responsible for the fence-like features by mounting electrodes inside the evaporation chamber. These electrodes deflect the ionized metal atoms and consequently the fence-like structures disappear. A similar deflection is achieved by the stray field of a magnet in close proximity to the sample. We note, however, that there are of course other mechanisms that can yield similar fence-like structures. Especially, if the deposition is carried out under an angle or if more isotropic deposition techniques such as sputtering are used. Then the resist sidewalls can be deposited with material similar to Fig.~2(e). Instead, for our ferromagnetic structures we observe that the overall extent of the fence-like structure scales with the degree of vapor ionization, which can explain a different amount of unintentional shadow evaporation in different evaporation systems. For example, the system used for the fabrication of the device shown in Fig.~2(a) operates at high e-beam currents, which results in a high density of primary and secondary electrons right above the crucible and, hence, a high chance of electron-impact ionization. The high currents are due to several factors: First, the system only operates at an acceleration voltage of \unit{4}{kV}, meaning that for the same power a much higher e-beam current must be applied. Second, no crucible liner is used for cobalt (cobalt alloys with refractory metals and tends to crack graphitic liners), which results in a high thermal conductance to the water-cooled hearth and therefore an increase in necessary e-beam power. Third, the size of the hearths in this system is quite small with a volume of only $\unit{2}{cm^3}$. This results in a steep temperature gradient. The area from which the metal evaporates is thus approximately equal to the area where the electron beam hits the surface of the metal. This increases the chance of electron-impact ionization even further as discussed in section~\ref{Partially_ionized_vapor}.

We also deposited ferromagnetic micromagnets, which are used for electric dipole spin resonance experiments \cite{Neumann2015,Struck2020} (see Fig.~2(c)). These micromagnets are made from cobalt and capped with a platinum layer in another e-beam evaporation system equipped with a larger crucible size ($\unit{7}{cm^3}$), with a higher acceleration voltage (\unit{8}{kV}), and with a higher deposition rate ($\unit{0.4}{nm/s}$). Accordingly, as long as other effects such as contamination of the evaporation material do not overcompensate the effects of the increased evaporation parameters, an overall lower degree of ionization is expected in this system. Indeed, the scanning electron micrograph of the micromagnet reveals a very thin halo-like shadow deposition around the whole structure in contrast to the very pronounced fence-like structure of the previously discussed cobalt contacts. Energy-dispersive X-ray spectroscopy (EDX) measurements on this structure confirm that the material of the halo-like shadow deposition is indeed the metal used during the evaporation process and not, e.g., cross-linked resist.

To completely avoid the shadow evaporation, the ionized part of the metal vapor must not reach the sample. One way to accomplish this is to create a high enough transverse electrostatic field between the crucible and the sample which deflects all charges. The most efficient way of creating such a field is to use two parallel electrodes in a plate capacitor geometry, so-called electron deflectors or ion collector plates.\cite{Nishio1993,Ohba1994,Pao1999,vanVeenhuizen2008,Baruah2014,Dikshit2007} The voltages, which are necessary to create high enough electric fields for the extraction of the ions, differ significantly as they depend on both the geometry of the electrodes and the charged vapor density due to space charge effects.

There is not always enough space in an evaporation system for placing two parallel electrodes, especially if the electrodes are retrofitted in an already existing system. One of the electrodes might block either the molecular beam of one of the sources or other components like shutters or quartz sensors. This is the case of the evaporation system which generates the shadow evaporation in Fig.~2(a). Nevertheless, we show that one large electrode is sufficient for avoiding the shadow evaporation as long as the sample can also be put on an electric potential. Accordingly, Fig.~3(a) depicts the schematic layout of the used evaporation system after inserting the electrode. As every nearby metallic surface functions as the counter part of this electrode, it is important to insert grounded plates between this electrode and the e-beam evaporator to minimize their interaction.

\begin{figure*}[t]
	\includegraphics{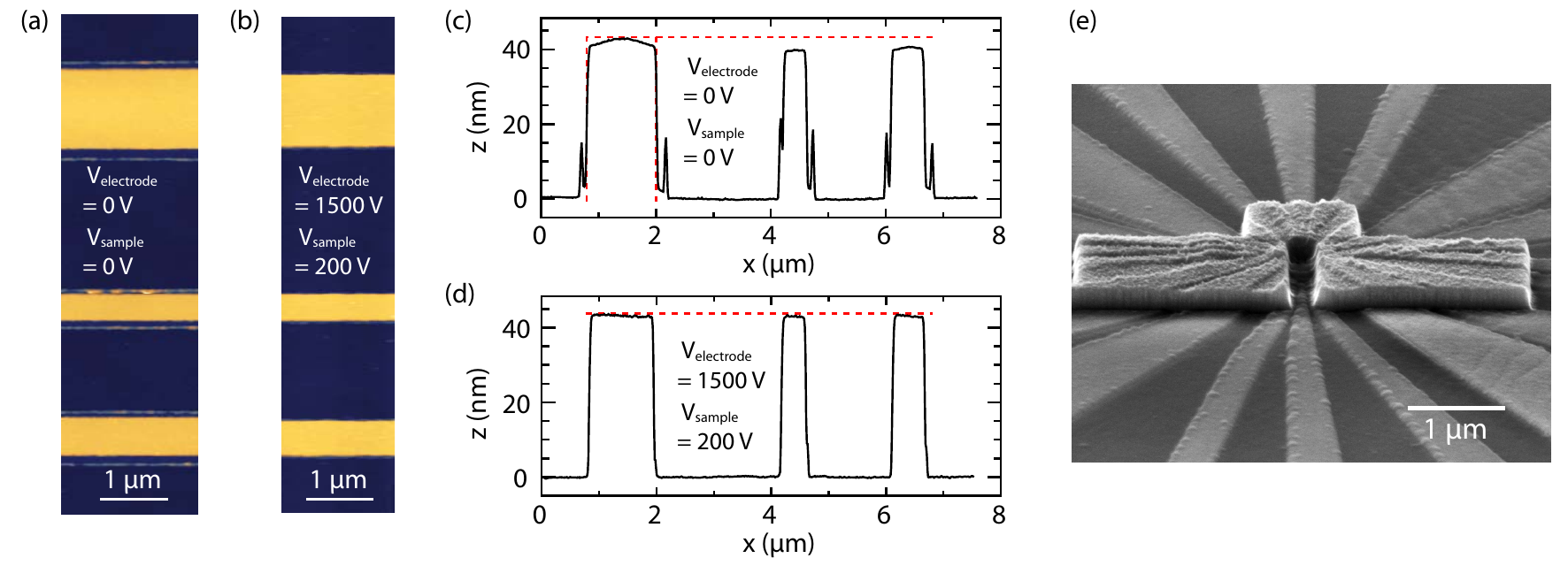}
	\caption{(a), (b) SFM images of ferromagnetic contacts deposited on a Si$^{++}$/SiO$_2$ substrate in case of (a) a grounded electrode and grounded sample (compare to Fig.~3(a)) and (b) with applied voltages of $V_\text{electrode}=\unit{1500}{V}$ and $V_\text{sample}=\unit{200}{V}$. Extracting the electrons from the metal vapor by the electrode and deflecting the positively ionized metal atoms from the sample results in the prevention of the unwanted shadow evaporation. (c), (d) Line-cuts of the SFM images shown in (a) and (b), respectively. (e) By increasing the deposition rate and by switching to another e-beam evaporation system which was equipped with a magnet to deflect charged particles away from the sample, the shadow evaporation also disappeared in case of the micromagnet (see highly defined sharp edges compared to Fig.~2(c)). The SEM image shows a micromagnet, which was deposited directly on top of gate electrodes.}
	\label{Fig4}
\end{figure*}

The amount of charges reaching the sample holder with the substrate (from here on, this whole assembly is denoted as 'sample' for the sake of simplicity) in case of a grounded electrode is shown in Fig.~3(b) as a function of the potential applied to the sample and varying e-beam currents. We note that the used evaporator does not measure the e-beam current directly and only has a nominal e-beam power given in a percentage reading. The current $I_\text{sample}$ flowing over the sample is measured by a source-meter unit (SMU), which simultaneously holds the sample to a fixed potential $V_\text{sample}$ against ground (see Fig.~3(a)). To avoid the shadow evaporation, a positive potential must be applied to the sample. This positive bias deflects positively charged metal ions from the substrate, but at the same time attracts electrons. The latter leads to a significant increase of the total current (see Fig.~3(b), plotted are the technical currents, therefore, a positive current means an electron flow from the sample over the SMU to ground).

For negative sample voltages the sample current shows a far less pronounced increase as a function of e-beam current (see Fig.~3(c), which depicts a zoom-in into Fig.~3(b)). We note that the used e-beam powers are below the threshold for evaporation. Therefore, the negative current (electrons flowing from ground to the sample) is not due to a compensation current of positively charged metal atoms. Rather, this current is due to both secondary electrons and photo-electrons created by high energy electrons and X-rays hitting the sample. The secondary electrons and photo-electrons are then repelled from the sample due to the negative potential compared to the grounded chamber wall.

Although a positive potential $V_\text{sample}$ to the sample is necessary to avoid the unintended shadow evaporation, the increased bombardment with electrons as seen in Fig.~3(b) results in one significant problem, which will be discussed in detail in section~\ref{Damage_Resist_Lift-Off_Problems}: The electrons can chemically alter the structure of a resist layer, which leads to problems in the lift-off. To prevent the electrons from reaching the sample, the electrode in Fig.~3(a), which was grounded so far, is now put to both positive and negative potentials $V_\text{electrode}$ to extract or repel the electrons. The results are shown in Fig.~3(d) in case of a constant deposition rate and a sample which is hold to ground potential by the SMU. The current $I_\text{electrode}$ flowing over the electrode to ground significantly increases for positive voltages, as the electrode extracts the electrons from the plasma. At the same time, the amount of electrons reaching the sample ($I_\text{sample}$) drops significantly.

\begin{figure*}[tb]
	\includegraphics{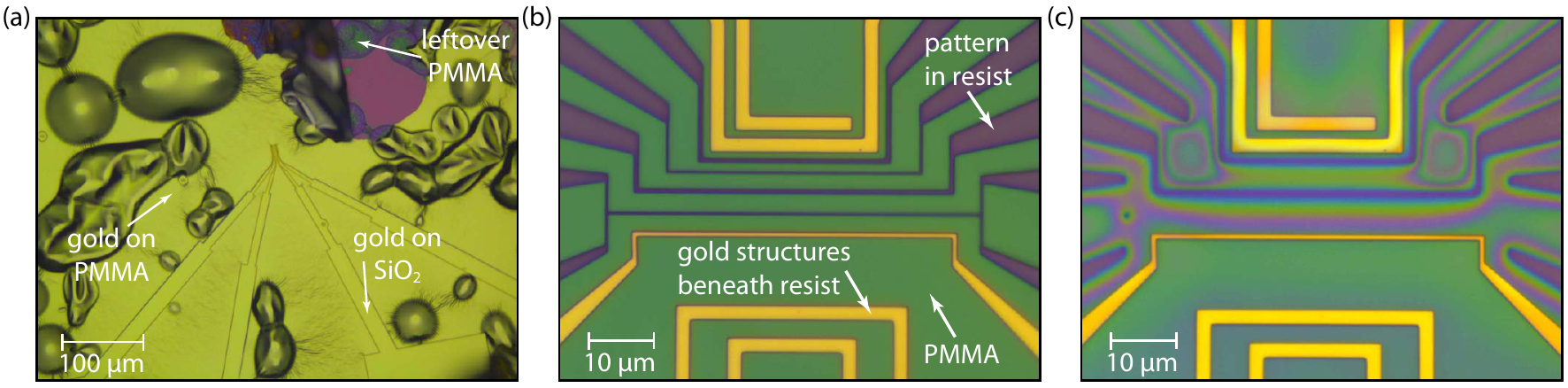}
	\caption{(a) Optical microscope image of a structure defined in a PMMA resist layer by electron beam lithography directly after the metallization process in an e-beam evaporator which had the problem of a high secondary electron emission. In places where the bubbles and blisters in the metal layer burst open, it is observed that the PMMA underneath these bubbles is partially gone. (b) A patterned structure written in a double layer resist layer spin coated on a Si$^{++}$/SiO$_2$ substrate already having gold structures. (c) Same sample as in (b) after heating the sample above the glass transition temperature of PMMA (around $\unit{105}{\degree C}$) which destroyed the smallest pattern in the resist.}
	\label{Fig5}
\end{figure*}

In case of negative voltages applied to the electrode (Fig.~3(d)), which lead to a repulsion of electrons away from the electrode towards the grounded chamber wall, the current over the sample initially drops before it rises again for larger negative voltages $V_\text{electrode}$. Figure~3(e) shows the current flowing over the sample in the same configuration as in Fig.~3(d) but over a wider voltage range (the continuous line is the measurement from Fig.~3(d)). For a voltage of around $V_\text{electrode}=\unit{-700}{V}$ the current reaches a maximum, which is higher than the current in case of a grounded electrode ($V_\text{electrode}=\unit{0}{V}$), before the current drops again for higher negative voltages. We attribute this maximum to two effects: On the one hand, secondary electrons and photo-electrons created at the electrode are accelerated towards the sample in case of a more negative potential of the electrode compared to the sample. On the other hand, the backscattered and secondary electrons from the e-beam evaporator have a different angular distribution compared to the neutral metal vapor, as the electrons are deflected by the same magnetic field, which focuses the incoming e-beam onto the surface of the crucible.\cite{Yamada2011} Therefore, applying a negative voltage to the electrode is most likely shifting the maximum of the angular distribution of backscattered and secondary electrons towards the sample. Overall, the data in Figs.~3(d) and 3(e) demonstrate that positive voltages applied to the electrode are more suitable to prevent electrons from reaching the sample.

Figure~3(f) shows the change of $I_\text{sample}$ as a function of the applied voltage $V_\text{sample}$ to the sample for different electrode potentials $V_\text{electrode}$ all at a constant deposition rate of the evaporated cobalt. For the grounded electrode ($V_\text{electrode}=\unit{0}{V}$, black curve), the strong increase in current towards higher sample voltages can be seen as discussed in Fig.~3(b) (the overall current is now higher as the e-beam power is increased for an actual evaporation process). But an increase of the electrode's voltage remarkably diminishes the amount of electrons reaching the sample.

We conclude that a positive potential applied to both the sample and the electrode is required to avoid shadow evaporation. This can be seen in the SFM data shown in Figs.~4(a) to 4(d). Here, similar ferromagnetic Co contacts were deposited onto Si$^{++}$/SiO$_2$ substrates as the one presented in Fig.~2(a). In case of Figs.~4(a) and 4(c) both the electrode in the evaporation chamber and the sample were grounded ($V_\text{electrode}=V_\text{sample}=\unit{0}{V}$), which again results in a pronounced shadow evaporation. On the other hand, Figs.~4(b) and 4(d) show the result of a deposition run where the electrode potential is set to $V_\text{electrode}=\unit{1500}{V}$ to extract electrons from the metal vapor. Additionally, the sample is put to $V_\text{sample}=\unit{200}{V}$ to deflect the ionized metal atoms. As a result, both the fence-like structure and the round edges of the contacts can be completely avoided. As an alternative to a deflection electrode a magnet can be placed in close proximity to the sample. The Lorentz force due to its stray field deflects both electrons and ionized metals.\cite{Knoch2020} We grew a Co micromagnet with the same geometry as the one shown in Fig.~2(c) in a third evaporation system equipped with such a deflection magnet. The result is shown in the scanning electron micrograph depicted in Fig.~4(e). Indeed the shadow evaporation observed in Fig.~2(c) is fully suppressed. Note that the micromagnet shown in Fig.~4(e) is grown on a thin metallic nano-gate structure fully covered by an insulating aluminum-oxide layer, which has no effect on the shadow evaporation.

\section{Damage of resist layer and lift-off problems caused by electrons}
\label{Damage_Resist_Lift-Off_Problems}
Another effect caused by charges emitted from an e-beam evaporator is cross-linking, blisters, and bubbles in a resist layer after e-beam evaporation. Figure~5(a) shows an optical microscope image of such an example. For this sample, first electron beam lithography was used to define micro- and nanostructures in a PMMA resist layer spin coated on top of a Si$^{++}$/SiO$_2$ substrate. After the development of the structure, it was metalized with \unit{5}{nm} Cr and \unit{50}{nm} Au in an e-beam evaporation chamber, which suffered from a high emission of secondary electrons. The image in Fig.~5(a) was recorded right after the metalization process and before lift-off. Therefore, the missing PMMA, which is observed in the area where some of the bubbles burst open and therefore removed the deposited metal layer on top, is thus not due to dissolving of the resist but rather results from a decomposition of the PMMA during metal evaporation.

\begin{figure*}[tb]
	\includegraphics{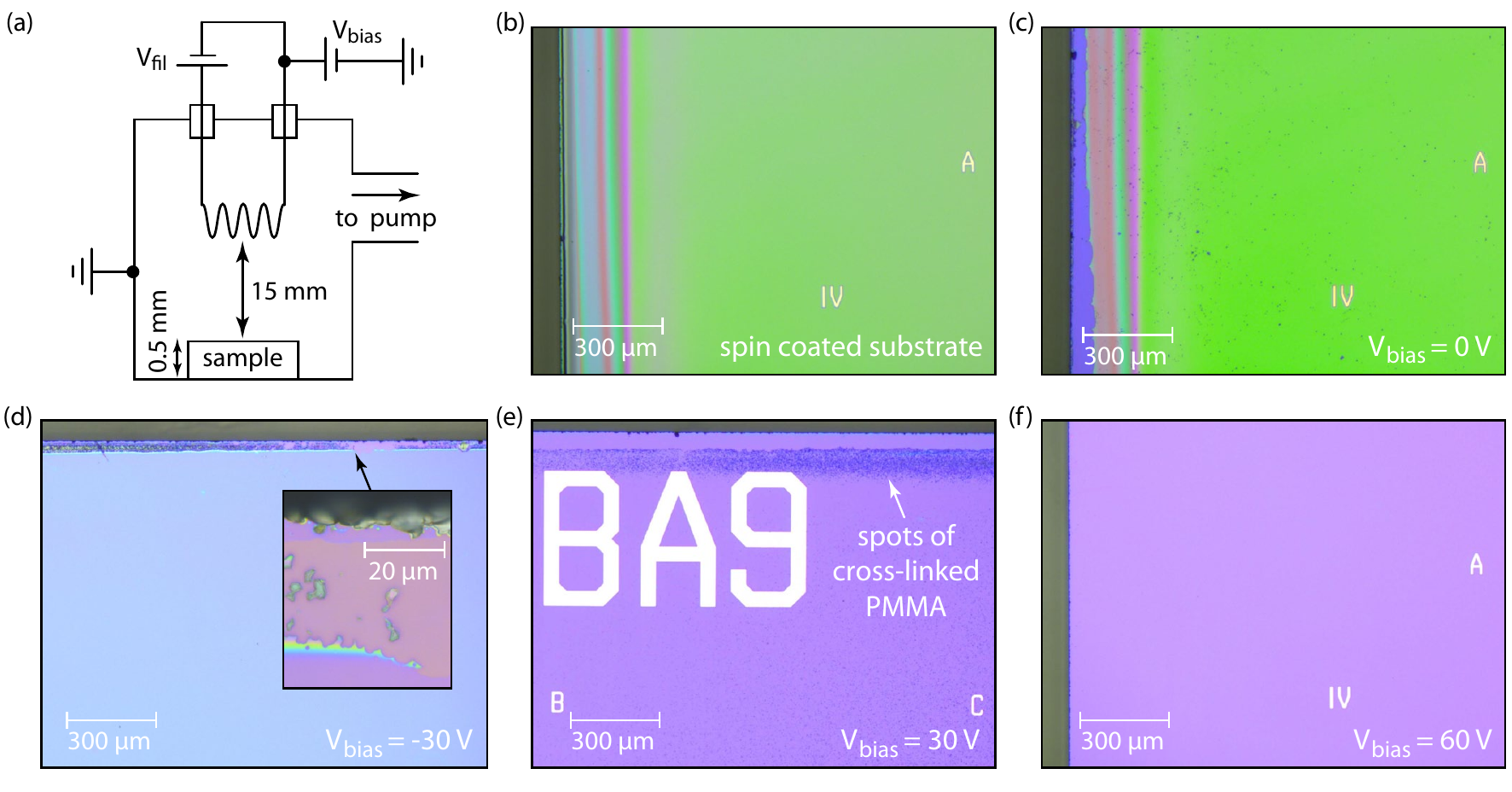}
	\caption{(a) Schematic layout of the vacuum chamber used to investigate the impact of low-energy electrons on a PMMA layer. Si$^{++}$/SiO$_2$ substrates with gold markers are spin coated with PMMA (see optical image in (b)) and are put directly beneath a filament. The filament assembly can be set to an arbitrary potential compared to the grounded chamber wall via $V_\text{bias}$. (c)-(f) Different spin coated substrates are exposed for the same amount of time (\unit{15}{s}) to the same filament power (\unit{45}{W}) but at different bias voltages $V_\text{bias}=\unit{-30...60}{V}$. The optical images were recorded after the substrates were put in an ultrasonic bath of acetone. Only for a bias voltage of $V_\text{bias}=\unit{60}{V}$ the PMMA can be removed completely by acetone, for all other voltages cross-linking occurs to a different extent. The samples (c) and (d) did not change during the acetone process showing that especially for (d) a significant amount of PMMA was etched away during the electron bombardment.}
	\label{Fig6}
\end{figure*}

A commonly encountered explanation for this problem is a too high temperature of the substrate during the deposition process. Although we do not exclude the possibility that excessive heat can be a problem for certain resists and deposition parameters, this is not the case in our study. First of all, the lift-off process of a sample as shown in Fig.~5(a) can be successful despite the severe damage of the resist. In such cases even nanometer-scale structures can be intact and well-defined. This observation contradicts a heat-induced damage of the resist, as small structures in the resist layer get destroyed quite easily as soon as the temperature rises above the glass transition temperature of the resist (around $\unit{105}{\degree C}$ for PMMA). This is demonstrated in Figs.~5(b) and 5(c), which show optical microscope images of nanostructures in a PMMA layer. Figure~5(b) shows this structure after the development, whereas the image in Fig.~5(c) was taken after heating beyond the glass transition temperature under vacuum conditions. Every structure with a width below \unit{1}{\micro m} is destroyed due to the thermal deformation of the resist. As nanostructures were intact in some of the samples showing the resist problems, we know that the temperature of the resist during the deposition was well below $\unit{100}{\degree C}$. But at these temperatures no thermal decomposition of PMMA should occur.

To exclude bubble generation by outgassing of volatile gases or solvents trapped within the resist, we baked the PMMA for \unit{2}{h} at \unit{180}{\degree C} before the transfer into the evaporation system. This temperature is much larger than the above discussed maximum resist temperature during deposition. But even under this condition, the damage to the resist and the creation of bubbles occurred. Therefore, the volatile gas species, which are responsible for the bubbles, have to be created during the evaporation process itself. Also the occurrence of cross-linking, i.e. chemically modified resist, which cannot be dissolved even in acetone,\cite{Teh2003} refutes thermal issues as we heat PMMA layers under vacuum conditions to temperatures of up to $\unit{500}{\degree C}$ but never observed thermally-induced cross-linking. On the other hand, cross-linking of PMMA is known to occur when the resist is exposed to irradiation by electrons, ions, or high energy photons.\cite{Lee1998,TruicaMarasescu2005} During this exposure to radiation, side groups in the PMMA chains are removed, which results in the creation of hydrogen and small hydrocarbon molecules.\cite{Hiraoka1977,Bermudez1999} Therefore, irradiation-induced decomposition of PMMA into volatile gas species can explain the creation of bubbles even in case of PMMA layers that were thoroughly degassed before the metalization process.

Note that even for a well-running evaporation system, resist problems as shown in Fig.~5(a) can occur suddenly after, e.g., a refilling of the crucibles and may vanish after the next replacement of the evaporation material or just after a re-adjustment of the e-beam. Therefore, the actual cause of resist problems can often be linked to contaminations of the evaporation material or a misalignment of the e-beam. Of all sources of particles and radiation emitted from an e-beam evaporator, which were discussed in section~\ref{Overview}, secondary electrons are most sensitive to such changes.

Low-energy electrons in the energy range below \unit{50}{eV} can indeed change the chemical structure of PMMA and other resists, e.g. via the process of dissociative electron attachment.\cite{Bermudez1999,Arumainayagam2010} To investigate the impact of these low energy electrons on our samples, we used a vacuum chamber in which the samples can be placed directly beneath a filament (Fig.~6(a)). The filament is connected to a floating power supply, which applies the voltage $V_\text{fil}$ necessary to drive the filament current. An additional power supply puts this filament assembly to an arbitrary potential $V_\text{bias}$ against the grounded chamber wall.

Figure~6(b) shows a Si$^{++}$/SiO$_2$ substrate which was spin coated with the same resist used for the sample shown in Fig.~5(a). Several of these substrates were then exposed for the same time (\unit{15}{s}) to the same filament power (\unit{45}{W}) at various bias voltages ranging from $V_\text{bias}=\unit{-30}{V}$ to \unit{60}{V}. As the exposure time and the electrical power of the filament was the same for all samples, the total exposure to infrared radiation should be the same. The only difference is the fact that thermionic electrons emitted from the hot filament are either accelerated from the filament towards the sample for $V_\text{bias}<\unit{0}{V}$ or reflected back towards the filament for $V_\text{bias}>\unit{0}{V}$. Optical microscope images were taken both right after this exposure and after an ultrasonic treatment of the substrates in an acetone bath. The images taken after the latter procedure are shown in Figs.~6(c)-(f).

We first discuss the case where the bias voltage was set to zero (Fig.~6(c)). Here, the acetone was not able to remove the resist layer on top of the substrate (for comparison: the color of a clean Si$^{++}$/SiO$_2$(\unit{285}{nm}) wafer as seen under the optical microscope is the same violet as in Fig.~6(f)). This means that the PMMA got cross-linked. For a bias voltage of \unit{-30}{V} (Fig.~6(d)) we note that the optical image right before the electron exposure looked comparable to Fig.~6(b), whereas the image right after the exposure looked identical to the one depicted in Fig.~6(d) after the acetone treatment. We conclude that the vast majority of the PMMA layer was etched away during the bombardment of electrons, which were accelerated towards the substrate at negative bias voltages. The leftover thin PMMA layer is completely cross-linked. In contrast, a positive bias voltage results in the deflection of thermionic electrons back to the filament. Therefore, for \unit{+30}{V} the PMMA layer was dissolved almost completely in acetone, only leaving behind smaller spots of cross-linked PMMA on the substrate (Fig.~6(e)). An important conclusion from this sample is the fact that even the smallest currents of low energy electrons have a significant detrimental effect on the PMMA layer. Because only an extremely small fraction of the thermionic electrons emitted from the filament can overcome the repulsive field caused by the potential difference of \unit{30}{V}. Eventually, at a bias voltage of \unit{+60}{V} the appearance of the PMMA layer did not change at all during exposure and it was completely dissolved in acetone, leaving behind a clean substrate (Fig.~6(f)). 

\begin{figure}[tb]
	\includegraphics[width=0.4\textwidth]{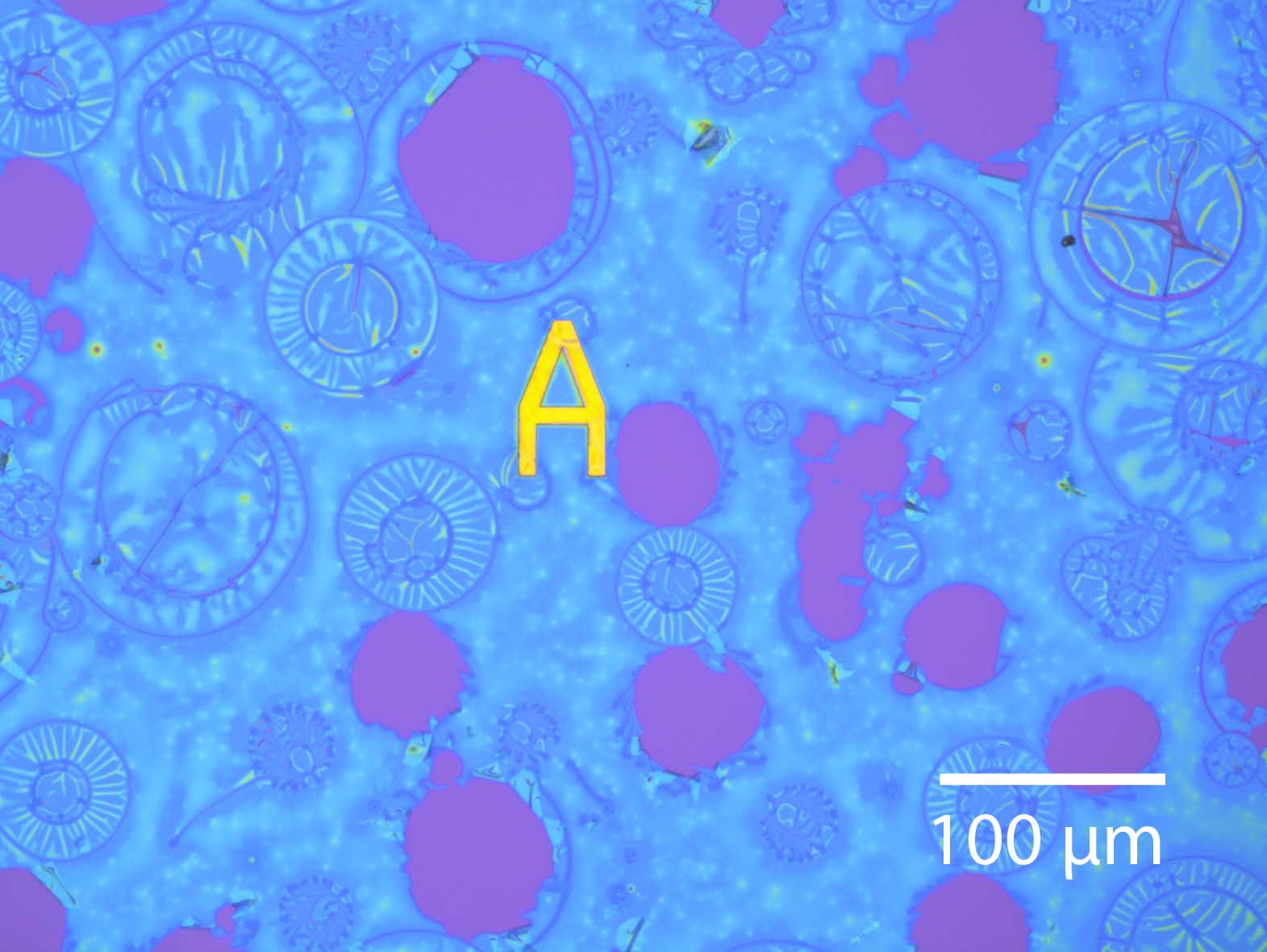}
	\caption{Optical image of a Si$^{++}$/SiO$_2$ substrate with a gold marker and spin coated PMMA layer, which was exposed two times to the electron bombardment described in Fig.~6 at a bias voltages of $V_\text{bias}=\unit{0}{V}$. Both the size and the shape of the bubbles and blisters are comparable to the sample with the resist problems in Fig.~5(a), which was metalized in an e-beam evaporator.}
	\label{Fig7}
\end{figure}

In a next step we altered the above procedure and exposed a substrate, which was spin coated with PMMA, two times to the radiation of the filament both at a bias voltage of $V_\text{bias}=\unit{0}{V}$: One shorter exposure (\unit{15}{s} at \unit{45}{W}) was followed by a longer exposure at slightly higher filament power (\unit{35}{s} at \unit{50}{W}). This procedure mimics the metalization process, during which we sequentially deposit two metals: First a thin metal layer of Ti or Cr, which is used as an adhesion layer, followed by a thicker layer of Au deposited at a higher rate motivating the higher filament power. The resulting optical image of the PMMA-covered substrate is shown in Fig.~7. We note that the observed bubbles and blisters have similar sizes and shapes compared to the sample with the resist problems shown in Fig.~5(a). Thus, the source of both resist damages is likely the same, i.e. it is caused by a bombardment of low-energy electrons.

The last argument that low-energy secondary electrons are the primary source of resist problems is the fact that we routinely get rid of these problems by following every procedure that can either reduce the emission of secondary electrons from the crucible or prevent them to reach the substrate: We check if there is a contamination of the evaporation material and replace it. If necessary,\cite{Cheng2010,Gonzalez2017,Baglin2000} we realign the electron-beam and change the deposition parameters,\cite{Seiler1983,Thomas1969,Kanaya1972} and we insert electrodes inside the chamber to extract the electrons. Alternatively, we install a magnet close to the sample in order to deflect charged particles away from the sample.\cite{Ohba1994,Nishio1993,Pao1999,vanVeenhuizen2008,Baruah2014,Dikshit2007,Knoch2020}

Finally, we would like to stress that resist problems are not solely linked to the condition of the evaporation system. Another important factor is, for example, the exact composition and type of resist, as the chemistry and especially the interaction with low energy electrons can differ significantly between different resists. Next to PMMA, we also investigated the photoresist AZ 5214 E and made similar deposition runs. We found that the AZ 5214 E resist was less prone to the creation of blisters and bubbles, but crosslinking and lift-off problems were still a severe issue, which could be overcome by the same procedures as described above.

\section{Troubleshooting}
\label{Troubleshooting}
As explained in the previous sections, possible problems during e-beam evaporation are related to various evaporation parameters (e.g. deposition rate, acceleration voltage, alignment of the e-beam), the evaporation material (e.g. exact kind of material and possible contaminations), the specific design of the e-beam evaporator (e.g. crucible size, used liners) and the design of the vacuum chamber (base pressure, residual gas composition, and especially the geometry of the chamber and the distance and angle between the crucible and the sample as the radiation emitted from the crucible has an angular distribution). Due to large chamber-to-chamber variations in all of these parameters, we note that strategies in minimizing evaporation-related problems may solve these issues in one system, but at the same time may have unsatisfactory results in other systems. 

\subsection{Workarounds to diminish the problems}
If an immediate opening of the vacuum chamber for a time-consuming troubleshooting is unfeasible, one of the following three workarounds might be applied to reduce the occurring problems:

1.) Higher deposition rates: We observe that both the resist issues and the shadow evaporation beneath the undercut of the resist layer get worse for lower deposition rates. At higher deposition rates bubbles and blister may still occur, but especially the cross-linking of the resist is significantly reduced. We attribute this to the explanation discussed in detail in section~\ref{Electromagnetic_radiation}: In a first order approximation we can assume that the temperature of the evaporation material increases linearly with the e-beam current. According to the Antoine equation, this results in a highly non-linear deposition rate as a function of e-beam current. On the other hand, the SEY should scale linearly with the e-beam current. Hence, for a given total thickness of the deposited material, the total amount of electrons reaching the resist can be significantly reduced by increasing the deposition rate, which results in a reduced overall deposition time (see similar argumentation in case of X-rays in Ref.~\onlinecite{Yamada2011}). However, a noteworthy drawback of higher deposition rates is a significant impact on the morphology of the deposited material, which e.g. can result in an increased surface roughness.\cite{AppliedSurfaceScience.115.3138,AppliedSurfaceScience.183.223229} Especially for structures in the lower nanometer range, e.g. for gates in qubit devices,\cite{Hollmann2020} we observe that an increasing grain size due to higher deposition rates limits the minimum achievable size of structures at which conductivity is still acceptable. The grain size will also give a fundamental limit to the size of gates which can be defined with a lift-off process. 

2.) Decreasing e-beam current by increasing acceleration voltage: This procedure is not available if the system is designed for a fixed acceleration voltage. In any case, it is important to realign the e-beam as soon as the acceleration voltage is changed. A potential drawback of an increased acceleration voltage might be possible damages due to the higher energy of backscattered electrons and X-rays.

3.) Focussing the e-beam: We observe that decreasing the visible spot size of the e-beam on the surface of the evaporation material often reduces deposition problems, as this focussing can reduce the volume of space right above the crucible in which e-beam and vapor cloud overlap.\cite{Bhatia1998}

\subsection{Identifying contaminations of evaporation material}
\label{IdentifyingContaminations}
Contaminations of the evaporation material might be the reason for sudden resist problems in an otherwise well-performing evaporation system. Although every measurement technique, which is capable of analyzing the chemical structure of a surface is suitable for identifying contaminations, we discuss two additional helpful methods in the following.

1.) Cathodoluminescence: In this process the transfer of energy from the e-beam to a material with a band gap excites valence band electrons over the band gap into a conduction band. The recombination of the created electron-hole pairs can cause the emission of photons within the visible spectrum.\cite{Garcia2010} As long as there is a direct line of view onto the evaporation material, the following quick test for metal surface contaminations can be done even in the operating system without the need to open the chamber. For this test, the power of the e-beam is slowly increased until the material just starts to evaporate. In case of a clean metal, it should only be observed how the metal starts to glow over its whole surface in a reddish to orange color due to thermal radiation at high e-beam currents. But if a greenish or bluish spot appears at low e-beam currents, even before the metal starts to glow due to thermal radiation, the metal surface must be covered with some material exhibiting a band gap. Freshly refilled evaporation materials e.g. may show cathodoluminescence shortly before the very first evaporation due to a native oxide layer.

2.) Scanning electron microscopy: If available, SEM can be a powerful tool to check if a putative resist problem results from an increased emission of secondary electrons by contaminations, as the imaging in an SEM is mostly accomplished by recording exactly such electrons.\cite{Seiler1983} A comparison of SEM images of a clean Au crucible with one covered by carbon contamination can e.g. be found in Ref.~\onlinecite{Cheng2010}.

\subsection{Sources of contamination}
In the following we list some general ideas on possible sources of contamination.

1.) Purity of the source material: To be on the safe side, only evaporation grade materials with low impurity levels should be used. These materials should be stored properly, so that e.g. metals prone to oxidation do not exhibit extensive oxide layers.

2.) Poor cleaning: This does not only include improper handling and cleaning techniques when replenishing the evaporation material, but also improper cleaning of both the vacuum chamber and the evaporator. For example, deposited materials from surfaces right above the evaporator should be removed as they may peel off and fall into the crucible. 

3.) Electron beam induced deposition (EBID): An electron beam within a vacuum chamber is known to interact with the residual gases inside the system. If there are hydrocarbons present (e.g. by an oil-sealed backing pump or by excessive cleaning with organic solvents without subsequent sufficiently long pumping and bake-out) the e-beam can deposit a layer of carbonized material.\cite{Stewart1934,vanDorp2008}

4.) Wrong crucible liner: Certain metals are highly reactive in their liquid phase and start to react with the crucible liner if the wrong liner material was chosen (manufacturers of e-beam liners normally provide guidelines and selection charts).

\subsection{Estimating resist temperature during deposition}
As we are not excluding the possibility that resist problems in other systems are due to thermal issues instead of electron bombardment, we summarize different ways to estimate the resist temperature during a metalization process.

1.) Glass transition temperature of a resist: As shown in Figs.~5(b) and 5(c), it is possible to determine if the temperature of the resist increased above the glass transition temperature by checking if sub-micron-sized structures in the resist layer are still intact after the evaporation process. Especially SFM or SEM imaging before and after the metalization process can give precise information about changes in the resist.

2.) Thermocouples: The measurement of the temperature of the sample or the sample holder by an attached thermocouple is a clean procedure even suitable for UHV chambers. The significant drawback here is the much higher heat capacity of the wires and the sample holder compared to the thin resist layer. Furthermore, the thermocouple will also have an overall better thermal coupling to components at room temperature than the resist layer. Therefore, the temperature measured with a thermocouple can be quite different to the actual resist temperature.

3.) Infrared (IR) thermometer: If the vacuum chamber can be equipped with a viewport, which has direct line-of-sight onto the substrate during the deposition process, an infrared thermometer can be used to probe the substrate temperature. However, special care has to be taken in matching the transmission spectrum of the viewport window to the specific IR sensor. Even in case of window materials with reasonable IR transmission like zinc selenide, a thoroughly calibration of the IR thermometer may be necessary to account for the still existing absorption.

\subsection{Reducing electron exposure to the sample}

There are several procedures for reducing the overall electron exposure of the sample during a metalization process.

1.) Realignment of the e-beam system: The secondary electron yield depends on the angle of incidence between e-beam and evaporation material.\cite{Seiler1983,Thomas1969,Kanaya1972} Furthermore, the angular distribution of backscattered and secondary electrons depends on the magnetic field, which focuses the incoming e-beam onto the surface of the crucible.\cite{Yamada2011} Therefore, the alignment of the e-beam should be in accordance to the manufacturer's specification. If the manual provides values of magnetic field strengths for specific positions of the evaporator, a verification of these values via a Hall sensor is advisable. Demagnetization of a permanent magnet may change the spatial distribution of the magnetic field. Finally, the interaction of the e-beam with the vapor above the crucible can vary with the focal point and spot diameter of the e-beam.\cite{Bhatia1998}

2.) Removing contaminated evaporation material: The secondary electron yield can significantly increase in the presence of contaminants, an oxide layer, or adsorbates, which cover the material.\cite{Gonzalez2017,Baglin2000} It has been reported that a sufficiently high carbon contamination of a gold crucible can even lead to a situation where the incoming e-beam is reflected or converted to secondary electrons to such a large extent that evaporation was no longer possible.\cite{Cheng2010}

3.) Checking the impedance of crucible to ground: Using an insulating e-beam liner may lead to a significant charging of the whole crucible, which may deflect electrons towards the sample. It is thus advisable to verify a low impedance connection between evaporation material and the common ground of the system. 

4.) Putting electrodes into the chamber: Electrodes can significantly diminish the amount of charges reaching the sample.\cite{Ohba1994,Nishio1993,Pao1999,vanVeenhuizen2008,Baruah2014} If the sample holder is connected to ground, it is advisable to put an electrode right next to the sample holder, parallel to the substrate. This electrode can then be used to estimate the amount of charges reaching the sample, which helps to correctly align the e-beam.

5.) Applying a magnetic field: Placing a permanent magnet into the chamber near the sample or the evaporation source can reduce the amount of electrons reaching the sample.\cite{Knoch2020} In general, a magnet closer to the evaporation source can be more efficient as even small deflection angles yield large overall deflections due to the travel distance to the sample. Due to the vicinity to the source, such magnets may need water cooling to prevent overheating. A readjustment of the e-beam might be necessary as the additional magnet may impact the magnetic focussing of the e-beam onto the evaporation source.

\section{Conclusion}
We demonstrated how the emission of electrons and ions during electron-beam physical vapor deposition leads to problems during micro- and nanofabrication processes. We proved that electrostatic deflection of incoming ionized metal vapor due to different potentials between the metal layer on top of the resist and the deposited structure can result in an unintentional shadow evaporation beneath the undercut of a resist layer. By inserting deflection electrodes inside the evaporation chamber or a magnet close to the sample, the shadow evaporation can be fully suppressed. Furthermore, we demonstrated how low-energy secondary electrons can cause cross-linking, blisters, and bubbles in a resist layer during the metalization process. We discussed and demonstrated different recipes to minimize and even solve this problem, which include the identification of contamination of the evaporation material, realignment of the e-beam, and particularly the insertion of deflection electrodes.

\begin{acknowledgments}
This project has received funding from the European Union's Horizon 2020 research and innovation programme under grant agreement No. 881603 (Graphene Flagship) and the Deutsche Forschungsgemeinschaft (DFG, German Research Foundation) under Germany's Excellence Strategy - Cluster of Excellence Matter and Light for Quantum Computing (ML4Q) EXC 2004/1 - 390534769, through DFG (BE 2441/9-1 and STA 1146/11-1), and by the Helmholtz Nano Facility (HNF)\cite{HNF} at the Forschungszentrum J\"ulich.
\end{acknowledgments}

\end{document}